\documentclass[onecolumn,english,smallextended]{svjour3}
\usepackage{amsmath}
\usepackage{graphicx}
\usepackage{amssymb}

\usepackage{babel}
\begin{document}

\title{How to include fermions into General relativity by exotic smoothness}

\author{Torsten Asselmeyer-Maluga \and Carl H. Brans}

\institute{T. Asselmeyer-Maluga \at German Aerospace Center (DLR), Berlin \\ \email{torsten.asselmeyer-maluga@dlr.de}
\and C.H. Brans \at Loyola University, New Orleans\\ http://www.loyno.edu/$\sim$brans \\ \email{brans@loyno.edu}
}

\date{Received: date / Accepted: date}
\maketitle
\begin{abstract}
This paper is two-fold. At first we will discuss the generation of
source terms in the Einstein-Hilbert action by using (topologically
complicated) compact 3-manifolds. There is a large class of compact
3-manifolds with boundary: a torus given as the complement of a (thickened)
knot admitting a hyperbolic geometry, denoted as hyperbolic knot complements
in the following. We will discuss the fermionic properties of this
class of 3-manifolds, i.e. we are able to identify a fermion with
a hyperbolic knot complement. Secondly we will construct a large class
of space-times, the exotic $\mathbb{R}^{4}$, containing this class
of 3-manifolds naturally. We begin with a topological trivial space,
the $\mathbb{R}^{4}$, and change only the differential structure
to obtain many nontrivial 3-manifolds. It is known for a long time
that exotic $\mathbb{R}^{4}$'s generate extra sources of gravity
(Brans conjecture) but here we will analyze the structure of these
source terms more carefully. Finally we will state that adding a hyperbolic
knot complement will result in the appearance of a fermion as source
term in the Einstein-Hilbert action.
\end{abstract}

\section{Introduction}

General relativity (GR) has changed our understanding of space-time.
In parallel, the appearance of quantum field theory (QFT) has modified
our view of particles, fields and the measurement process. The usual
approach for the unification of QFT and GR, to a theory of quantum
gravity, starts with a proposal to quantize GR and its underlying
structure, space-time. There is a unique opinion in the community
about the relation between geometry and quantum theory: The geometry
as used in GR is classical and should emerge from a quantum gravity
in the limit (Planck's constant tends to zero). Most theories went
a step further and try to get a space-time from quantum theory. Then,
the model of a smooth manifold is not suitable to describe quantum
gravity. There is no evidence for discrete space-time structure or
higher dimensions in current experiments. Therefore, we will consider
a smooth 4-manifold as a model to describe the space-time in classical
and quantum gravity. This view is not in conflict with quantized areas
and volumes, see the example of Mostow rigidity (in subsection \ref{sub:Fermions-as-knot-complement}).
But then one has the problem to represent QFT by geometric methods
(submanifolds for particles or fields etc.) as well to quantize GR.
Here, the exotic smoothness structure of 4-manifolds can help to find
a way. A lot of work was done in the last decades to fulfill this
goal. It starts with the work of Brans and Randall \cite{BraRan:93}
and of Brans alone \cite{Bra:94a,Bra:94b,Bra:99} where the special
situation in exotic 4-manifolds (in particular the exotic $\mathbb{R}^{4}$)
was explained. One main result of this time was the {\em Brans conjecture}:
exotic smoothness can serve as an additional source of gravity. It
was confirmed for compact manifolds by Asselmeyer \cite{Ass:96} and
for the exotic $\mathbb{R}^{4}$ by S{\l}adkowski \cite{Sla:99,Sladkowski2001}.
But this conjecture was extended in \cite{AssBra:2002} to conjecture
the generation for all forms of known energy, especially dark matter
and dark energy. For dark energy we were partly successful in \cite{AsselmeyerKrol2014}
where we calculated the expectation value of an embedded surface.
This value showed an inflationary behavior and we were also able to
calculate a cosmological constant having a realistic value (in agreement
with the Planck satellite results). 

The inclusion of QFT was also another goal of our approach. We showed
\cite{AsselmeyerKrol2009} that an exotic 4-manifold (and therefore
the space-time) has a complicated foliation. Using noncommutative
geometry, we were able to study these foliations and got relations
to QFT. For instance, the von Neumann algebra of a codimension-1 foliation
of an exotic $\mathbb{R}^{4}$ must contain a factor of type $I\! I\! I_{1}$
used in local algebraic QFT to describe the vacuum\cite{AsselmeyerKrol2010,AsselmeyerKrol2011a,AsselmeyerKrol2011d}.
But why is an exotic 4-manifold so complicated? As an example let
us consider the exotic $S^{3}\times\mathbb{R}$. Clearly, there is
always a topologically embedded 3-sphere but there is no smoothly
embedded one. Let us assume the well-known hyperbolic metric of the
space-time $S^{3}\times\mathbb{R}$ using the trivial foliation into
leafs $S^{3}\times\left\{ t\right\} $ for all $t\in\mathbb{R}$.
Now we demand the exotic smoothness structure at the same time. Then
we will get only topologically embedded 3-spheres, the leafs $S^{3}\times\left\{ t\right\} $
(otherwise one obtains the standard smoothness structure, see \cite{Chernov2012}
for instance). These topologically embedded 3-spheres are also known
as wild 3-spheres. In \cite{AsselmeyerKrol2011c}, we presented a
relation to quantum D-branes. Finally we proved in \cite{AsselmeyerKrol2013}
that the deformation quantization of a tame embedding (the usual embedding)
is a wild embedding. Furthermore we obtained a geometric interpretation
of quantum states: wildly embedded submanifolds are quantum states.
Importantly, this construction depends essentially on the continuum,
wildly embedded submanifolds are always infinite triangulations. This
approach opens a way to quantize a theory using geometric methods.

For a special class of compact 4-manifolds we showed in \cite{AsselmeyerRose2012}
that exotic smoothness can generate fermions and gauge fields using
the so-called knot surgery of Fintushel and Stern \cite{FinSte:98}.
Here, the knot is directly related to the appearance of an exotic
smoothness structure, i.e. for two knots with different Alexander
polynomials (a knot invariant) one obtains non-diffeomorphic 4-manifolds.
From the physics point of view, the knot is somehow related to the
fermions (and gauge fields for complicated knots). Therefore, one
obtains a fixed configuration of fermions for every exotic 4-manifold
(using Fintushel-Stern knot surgery) or the number of fermions is
conserved. But in QFT, one is faced with the problem to have a variable
number of particles. This concept cannot be realized by using a fixed
configuration of fermions like in Fintushel-Stern knot surgery. Instead
one needs a more flexible exotic 4-manifold with a variety of submanifolds
which is related to the exotic smoothness structure. In this paper
we will present an approach using the exotic $\mathbb{R}^{4}$ which
will present a theory with variable particle number in difference
to \cite{AsselmeyerRose2012}. 

The results of this paper are two-fold: at first we will show how
to generated fermions from hyperbolic 3-manifolds and secondly we
will argue that the exotic $\mathbb{R}^{4}$ contains 3-manifolds
which can be interpreted as fermions. The special role of the exotic
$\mathbb{R}^{4}$ is given by the fact (for all known exotic $\mathbb{R}^{4}$)
that every neighborhood of a compact subset in the exotic $\mathbb{R}^{4}$
is surrounded by a compact 3-manifold (not homeomorphic to the 3-sphere)
but cannot be surrounded by a (smoothly embedded) 3-sphere. Therefore
we obtain always a non-trivial 3-manifold from an exotic $\mathbb{R}^{4}$
whereas for the standard $\mathbb{R}^{4}$ one can always choose a
neighborhood which is surrounded by a 3-sphere. But this non-trivial
3-manifold in the exotic $\mathbb{R}^{4}$ is not uniquely determined,
it depends on the representation of the exotic $\mathbb{R}^{4}$ and
on the choice of the neighborhood. At this point we obtain a non-trivial
3-manifold which is not uniquely determined in contrast to \cite{AsselmeyerRose2012}.
Secondly, the Fintushel-Stern knot surgery in \cite{AsselmeyerRose2012}
uses directly the knot complement and the Dirac action follows by
a special choice of a surface (using the Weierstrass representation).
In this paper we will consider directly the embedding of the 3-manifold
into the 4-manifold to get the Dirac action on the 3-manifold as well
as an extension to the Dirac action on the 4-manifold. Then the 3-manifold
is represented by ($0-$framed) surgery along the $n$th (untwisted)
Whitehead double of some knot. The choice of the level of the Whitehead
double is one freedom but there is more room for ambiguities. By this
method one obtains a decomposition of the 3-manifold into knot complements
again (see subsection \ref{sub:Fermions-as-knot-complement})  but
the representation of the 3-manifold by this knot is not unique. Therefore
the decomposition of the 3-manifold can be changed by some operations
(usually called Kirby calculus, see \cite{GomSti:97}). One can interpret
this behavior in QFT where a fermion is surrounded by a 'cloud' of
virtual particles. For the exotic $\mathbb{R}^{4}$, we will obtain
even this picture of non-constant particle numbers like in QFT. In
our discussion above, we mixed the words fermion field, fermion and
particle but we have to be more careful. We obtain the Dirac action
from the embedding of the 3-manifold or the fermion field (fulfilling
the Dirac equation) describes the embedding directly. But a part of
this 3-manifold, the hyperbolic knot complement, has properties of
a fermion, a particle of spin$\frac{1}{2}$. In the QFT picture, the
excitations of the field are the particles. Then in our picture, the
properties of the fermion field are given by the particular embedding
and we will obtain the Dirac action for the general case. Now following
the philosophy of QFT, knot complements can be seen as a kind of excitation,
i.e. there are different realizations of the same 3-manifold by decompositions
using knot complements. We think that only very general properties
of the knot (used to get the knot complement) are connected with particle
properties like charge or mass. Whereas the dynamics are related to
the geometric properties of the embedding. In particular, the mean
curvature of the embedding is the eigenvalue of the 3-dimensional
Dirac operator (determining the 3-momentum, see equation (\ref{eq:Dirac3D-mean-curvature})
below). In our forthcoming work we will clarify this point of view
more deeply. 

In \cite{AsselmeyerRose2012} we obtained a complete picture of known
matter: fermions as hyperbolic knot complements and gauge fields as
torus bundles but we got only the action functionals. From the physics
point of view, a fermion is a particle of spin $\frac{1}{2}$ (or
$\frac{n}{2}$ in general) whose dynamics are described by the Dirac
equation (or Pauli-Fierz equation in general) and they are given by
the state equation $p=0$ (non-contractable matter) in the cosmological
context. We will discuss all these properties in section \ref{sec:The-physical-interpretation}
for our example of an exotic $\mathbb{R}^{4}$ but these properties
will go over to the example in \cite{AsselmeyerRose2012}. Furthermore
the relation to quantum gravity has to be understand more completely.
First signs of a relation can be found in \cite{Duston2010,Ass2010}
or by using string theory \cite{AssKrol2010ICM}. The results of this
paper seem to suggest that an exotic $\mathbb{R}^{4}$ does not fix
the concrete form of matter (fermions, gauge fields) but can fix the
rules how matter can change into each other. It agrees with a philosophical
interpretation of QFT where a particle is represented by a bundle
of properties (Dispositional Trope Ontology, see \cite{QFTinterpretation2012}).
Further work is necessary to support this conjecture.

Here is the plan of our paper. In the next section we will discuss
the different constructions of exotic $\mathbb{R}^{4}$'s (large and
small). But we will also describe the common property of all known
examples: there is a compact subset in any known exotic $\mathbb{R}^{4}$
which cannot be surrounded by a 3-sphere. Based on this property we
will also introduce the (Euclidean) Einstein-Hilbert action with boundary
term. In section \ref{sec:Dirac-action-3MF} we will describe our
main method to determine the boundary term: the embedding of the 3-manifold
can be described by spinors so that the boundary term of the Einstein-Hilbert
action is the Dirac action for this spinor. With some effort one can
extend this action from the boundary to some part of the space-time.
But more importantly, the boundary can be decomposed into knot complements
and we are able to interpret the knot complements admitting hyperbolic
geometry as fermions in section \ref{sec:The-physical-interpretation}.
In section \ref{sec:The-Brans-conjecture} we will discuss the Brans
conjecture, i.e. the generation of source terms in General Relativity
by using exotic smoothness. Our method can be also used to generate
fermions by adding non-trivial 3-manifolds as shown in section \ref{sec:Outlook}.
We will place special emphasis on spin networks and spin foams. Finally
we will summarize the results. Three appendices replenish our approach
with some technical details.

Finally we will state the main result of our paper: \emph{The matter
content of the universe can be interpreted as being located on non-trivial
3-manifolds, which are represented by hyperbolic knot complements
and graph manifolds.} \emph{Then in the standard $\mathbb{R}^{4}$,
we can always choose a 3-sphere to separate a compact subset from
infinity or expressed in physics language, we always obtain an empty
cosmos. In contrast, an exotic $\mathbb{R}^{4}$does not allow to
embed a (smooth) 3-sphere separating a compact subset from infinity.
But there is a non-trivial 3-manifold for separating this compact
subset. Therefore, one gets a cosmos (the non-trivial 3-manifold)
filled with matter. If one interprets one direction as time (for instance
a radial coordinate so that the compact subset has fixed size), then
matter does not exists for all times (only outside of the compact
subset). }

\section{Construction of exotic $\mathbb{R}^{4}$}

Our model of space-time is the non-compact topological $\mathbb{R}^{4}$.
The results can be easily generalized for other cases such as $S^{3}\times\mathbb{R}$.
In this section we will give some information about the construction
of exotic $\mathbb{R}^{4}$. The existence of a smooth embedding $R^{4}\to S^{4}$
of the exotic $\mathbb{R}^{4}$ into the 4-sphere splits all exotic
$\mathbb{R}^{4}$ into two classes, large (no embedding) or small.
We recommend the books \cite{Asselmeyer2007} (toward physical applications
of exotic smoothness), \cite{Scorpan2005} (for an overview of exotic
manifolds) and \cite{GomSti:97} (for the construction of exotic 4-manifolds).

\subsection{Preliminaries: Slice and non-slice knots}

\begin{figure}
\begin{center}\includegraphics[scale=0.7]{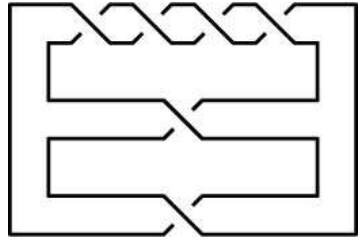}\end{center}

\caption{a smoothly slice knot: Stevedore's knot $6_{1}$\label{fig:Stevedore-knot-6_1}}
\end{figure}

At first we start with some definitions from knot theory. A (smooth)
knot $K$ is a smooth embedding $S^{1}\to S^{3}$. Furthermore, the
$n$-disk is denoted by $D^{n}$ with $\partial D^{n}=S^{n-1}$. \begin{definition}
\textbf{Smoothly Slice Knot}: A knot in $\partial D^{4}=S^{3}$ is
smoothly slice if there exists a two-disk $D^{2}$ smoothly embedded
in $D^{4}$ such that the image of $\partial D^{2}=S^{1}$ is $K$.
\end{definition} An example of a smoothly slice knot is the so-called
Stevedore's Knot (in Rolfsen notation $6_{1}$, see Fig. \ref{fig:Stevedore-knot-6_1}).

\begin{figure}
\begin{center}\includegraphics[angle=90,scale=0.8]{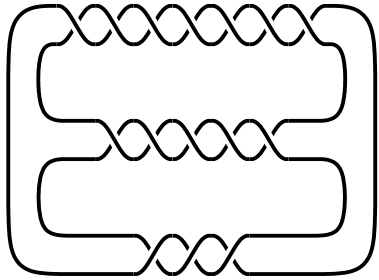}\end{center}

\caption{topological, non-smoothly slice knot: pretzel knot $(-3,5,7)$\label{fig:pretzel-knot-3-5-7}}
\end{figure}

\begin{definition} \textbf{Flat Topological Embedding}: Let $X$
be a topological manifold of dimension $n$ and $Y$ a topological
manifold of dimension $m$ where $n<m$. A topological embedding $\rho:X\to Y$
is flat if it extends to a topological embedding $\rho:X\times D^{m-n}\to Y$.

\textbf{Topologically Slice Knot}: A knot $K$ in $\partial D^{4}$
is topologically slice if there exists a two-disk $D^{2}$ flatly
topologically embedded in $D^{4}$ such that the image of $\partial D^{2}$
is $K$. \end{definition} Here we remark that the flatness condition
is essential. Any knot $K\subset S^{3}$ is the boundary of a disc
$D^{2}$ embedded in $D^{4}$, which can be seen by taking the cone
over the knot. But the vertex of the cone is a non-flat point (the
knot is crashed to a point). The difference between the smooth and
the flat topological embedding is the key for the following discussion.
Examples of slice knots were known for a long time. Clearly every
smoothly slice knot is also a topologically slice knot but whether
the reverse implication is true was not known until the work of Donaldson.
But deep results from 4-manifold topology gave a negative answer:
there are topologically slice knots which are not smoothly slice.
An example is the pretzel knot $(-3,5,7)$ (see Fig. \ref{fig:pretzel-knot-3-5-7}).

In \cite{Fre:82a}, Freedman gave a topological criteria for topological
sliceness: if the Alexander polynomial $\triangle_{K}(t)$ (the best
known knot invariant, see \cite{Rol:76}) of the knot $K$ is trivial,
$\triangle_{K}(t)=1$, then the knot $K$ is topologically slice.
But the converse is wrong in general. An example how to measure the
smooth sliceness is given by the smooth 4-genus $g_{4}(K)$ of the
knot $K$, i.e. the minimal genus of a surface $F$ smoothly embedded
in $D^{4}$ with boundary $\partial F=K$ the knot. Therefore, if
the smooth 4-genus vanishes $g_{4}(K)=0$ then the knot $K$ bounds
a 2-disk $D^{2}$ (surface of genus $0$) given by the smooth embedding
$D^{2}\to D^{4}$ so that the image of $\partial D^{2}\to\partial D^{4}$
is the knot $K$.

\subsection{Large exotic $\mathbb{R}^{4}$ and non-slice knots\label{sub:Large-exotic-R4}}

Large exotic $\mathbb{R}^{4}$ can be constructed by using the failure
to arbitrarily split a compact, simple-connected 4-manifold. For every
topological 4-manifold one knows how to split this manifold \emph{topologically}
into simpler pieces using the work of Freedman \cite{Fre:82}. But
as shown by Donaldson \cite{Don:83}, some of these 4-manifolds do
not exist as smooth 4-manifolds. This contradiction between the continuous
and the smooth case produces the first examples of exotic $\mathbb{R}^{4}$\cite{Gom:83}.
Unfortunately, the construction method is rather indirect and therefore
useless for applications of the exotic $\mathbb{R}^{4}$ in physics.
But as pointed out by Gompf (see \cite{Gom:85} or \cite{GomSti:97}
Exercise 9.4.23 on p. 377ff and its solution on p. 522ff), large exotic
$\mathbb{R}^{4}$ can be also constructed by using smoothly non-slice
but topologically slice knots. In the following we will use the notation:
$\mathbf{R^{4}}$ for the standard $\mathbb{R}^{4}$ and $R^{4}$
for the exotic $\mathbb{R}^{4}$, $\mathbb{R}^{4}$ will denote the
topological structure.

Let $K$ be a knot in $\partial D^{4}$ and $X_{K}$ the two-handlebody
obtained by attaching a two-handle to $D^{4}$ along $K$ with framing
$0$. That means: one has a two-handle $D^{2}\times D^{2}$ which
is glued to the 0-handle $D^{4}$ along its boundary using a map $f:\partial D^{2}\times D^{2}\to\partial D^{4}$
so that $f(.\,,\, x)=K\times x\subset S^{3}=\partial D^{4}$ for all
$x\in D^{2}$ (or the image $im(f)=K\times D^{2}$ is the solid knotted
torus). Let $\rho:X_{K}\to\mathbf{R}^{4}$ be a flat topological embedding
($K$ is topologically slice). For $K$ a smoothly non-slice knot,
the open 4-manifold 
\begin{equation}
R^{4}=\left(\mathbf{R}^{4}\setminus int\rho(X_{K})\right)\cup_{\partial X_{K}}X_{K}\label{eq:decomposition-large-exotic-R4}
\end{equation}
where $int\rho(X_{K})$ is the interior of $\rho(X_{K})$, is homeomorphic
but non-diffeomorphic to $\mathbf{R}^{4}$ with the standard smoothness
structure (both pieces are glued along the common boundary $\partial X_{K}$).
Importantly, the first term $\mathbf{R}^{4}\setminus int\rho(X_{K})$
in (\ref{eq:decomposition-large-exotic-R4}) has initially not a smooth
boundary. Then the smoothing of this boundary is rather complicated
(see chapter 8 in \cite{FreQui:90} or Theorem 9.4.22 of \cite{GomSti:97}). 

The proof of this fact ($R^{4}$ is exotic) is given by contradiction,
i.e. let us assume $R^{4}$ is diffeomorphic to $\mathbf{R}^{4}$.
Thus, there exists a diffeomorphism $R^{4}\to\mathbf{R}^{4}$. The
restriction of this diffeomorphism to $X_{K}$ is a smooth embedding
$X_{K}\hookrightarrow\mathbf{R}^{4}$. However, such a smooth embedding
exists if and only if $K$ is smoothly slice (see \cite{GomSti:97}).
But, by hypothesis, $K$ is not smoothly slice. Thus by contradiction,
there exists no diffeomorphism $R^{4}\to\mathbf{R}^{4}$ and $R^{4}$
is exotic, homeomorphic but not diffeomorphic to $\mathbf{R}^{4}$.
Finally, we have to prove that $R^{4}$ is large. $X_{K}$, by construction,
is compact and a smooth submanifold of $R^{4}$. By hypothesis, $K$
is not smoothly slice and therefore $X_{K}$ can not smoothly embed
in $\mathbf{R}^{4}$. Or, $R^{4}$ is a large exotic $\mathbb{R}^{4}$.

\subsection{Small exotic $\mathbb{R}^{4}$ and Casson handles\label{sub:Small-exotic-R4}}

Small exotic $\mathbb{R}^{4}$'s are again the result of anomalous
smoothness in 4-dimensional topology but of a different kind than
for large exotic $\mathbb{R}^{4}$'s. In 4-manifold topology \cite{Fre:82},
a homotopy-equivalence between two compact, closed, simply-connected
4-manifolds implies a homeomorphism between them (a so-called h-cobordism).
But Donaldson \cite{Don:87} provided the first smooth counterexample,
i.e. both manifolds are generally not diffeomorphic to each other.
The failure can be localized in some contractible submanifold (Akbulut
cork) so that an open neighborhood of this submanifold is a small
exotic $\mathbb{R}^{4}$. The whole procedure implies that this exotic
$\mathbb{R}^{4}$ can be embedded in the 4-sphere $S^{4}$.

The idea of the construction is simply given by the fact that every
such smooth h-cobordism between non-diffeomorphic 4-manifolds can
be written as a product cobordism except for a compact contractible
sub-h-cobordism $V$, the Akbulut cork. An open subset $U\subset V$
homeomorphic to $[0,1]\times{{\mathbb{R}}^{4}}$ is the corresponding
sub-h-cobordism between two exotic ${{\mathbb{R}}^{4}}$'s. These
exotic ${{\mathbb{R}}^{4}}$'s are called ribbon ${{\mathbb{R}}^{4}}$'s.
They have the important property of being diffeomorphic to open subsets
of the standard ${{\mathbb{R}}^{4}}$. To be more precise, consider
a pair $(X_{+},X_{-})$ of homeomorphic, smooth, closed, simply-connected
4-manifolds.

\begin{theorem}\emph{ }(\cite{GomSti:97} Theorem 9.3.1) Let $W$
be a smooth h-cobordism between closed, simply connected 4-manifolds
$X_{-}$ and $X_{+}$. Then there is an open subset $U\subset W$
homeomorphic to $[0,1]\times{{\mathbb{R}}^{4}}$ with a compact subset
$C\subset U$ such that the pair $(W\setminus C,U\setminus C)$ is
diffeomorphic to a product $[0,1]\times(X_{-}\setminus C,U\cap X_{-}\setminus C)$.
The subsets $R_{\pm}=U\cap X_{\pm}$ (homeomorphic to ${{\mathbb{R}}^{4}}$)
are diffeomorphic to open subsets of ${{\mathbb{R}}^{4}}$. If $X_{-}$
and $X_{+}$ are not diffeomorphic, then there is no smooth 4-ball
in $R_{\pm}$ containing the compact set $Y_{\pm}=C\cap R_{\pm}$,
so both $R_{\pm}$ are exotic ${{\mathbb{R}}^{4}}$'s.\emph{ }\end{theorem}

Thus, $R_{-}$ lies in a compact set, i.e. a 4-sphere or $R_{-}$
is a small exotic $\mathbb{R}^{4}$. In \cite{DeMichFreedman1992}
Freedman and DeMichelis constructed also a continuous family of small
exotic $\mathbb{R}^{4}$. Now we are ready to discuss the decomposition
of a small exotic $\mathbb{R}^{4}$ by Bizaca and Gompf \cite{BizGom:96}
by using special pieces, the handles forming a handle body. Every
4-manifold can be decomposed (seen as handle body) using standard
pieces such as $D^{k}\times D^{4-k}$, the so-called $k$-handle attached
along $\partial D^{k}\times D^{4-k}$ to the boundary $S^{3}=\partial D^{4}$
of a $0-$handle $D^{0}\times D^{4}=D^{4}$. The construction of the
handle body can be divided into two parts. The first part is the manifold
$Y_{-}$ in the theorem above, whereas the second part is the Casson
handle $CH$ which will be considered now.

Let us start with the basic construction of the Casson handle $CH$.
Let $M$ be a smooth, compact, simple-connected 4-manifold and $f:D^{2}\to M$
a (codimension-2) mapping. By using diffeomorphisms of $D^{2}$ and
$M$, one can deform the mapping $f$ to get an immersion (i.e. injective
differential) generically with only double points (i.e. $\#|f^{-1}(f(x))|=2$)
as singularities \cite{GolGui:73}. But to incorporate the generic
location of the disk, one is rather interested in the mapping of a
2-handle $D^{2}\times D^{2}$ induced by $f\times id:D^{2}\times D^{2}\to M$
from $f$. Then every double point (or self-intersection) of $f(D^{2})$
leads to self-plumbings of the 2-handle $D^{2}\times D^{2}$. A self-plumbing
is an identification of $D_{0}^{2}\times D^{2}$ with $D_{1}^{2}\times D^{2}$
where $D_{0}^{2},D_{1}^{2}\subset D^{2}$ are disjoint sub-disks of
the first factor disk%
\footnote{In complex coordinates the plumbing may be written as $(z,w)\mapsto(w,z)$
or $(z,w)\mapsto(\bar{w},\bar{z})$ creating either a positive or
negative (respectively) double point on the disk $D^{2}\times0$ (the
core).%
}. Consider the pair $(D^{2}\times D^{2},\partial D^{2}\times D^{2})$
and produce finitely many self-plumbings away from the attaching region
$\partial D^{2}\times D^{2}$ to get a kinky handle $(k,\partial^{-}k)$
where $\partial^{-}k$ denotes the attaching region of the kinky handle.
A kinky handle $(k,\partial^{-}k)$ is a one-stage tower $(T_{1},\partial^{-}T_{1})$
and an $(n+1)$-stage tower $(T_{n+1},\partial^{-}T_{n+1})$ is an
$n$-stage tower union of kinky handles $\bigcup_{\ell=1}^{n}(T_{\ell},\partial^{-}T_{\ell})$
where two towers are attached along $\partial^{-}T_{\ell}$. Let $T_{n}^{-}$
be $(\mbox{int}T_{n})\cup\partial^{-}T_{n}$ and the Casson handle
\[
CH=\bigcup_{\ell=0}T_{\ell}^{-}
\]
is the union of towers (with direct limit topology induced from the
inclusions $T_{n}\hookrightarrow T_{n+1}$).

The main idea of the construction above is very simple: an immersed
disk $D_{1}$ (disk with self-intersections) can be deformed into
an embedded disk $D_{\infty}$ (disk without self-intersections) by
sliding one part of the immersed disk $D_{1}$ along another disks
$D_{2}$ to kill the self-intersections (one disk for every self-intersection).
Unfortunately the disks $D_{2}$ can be immersed only. But the immersion
can be deformed to an embedding by disks $D_{3}$ etc. In the limit
of this process one ''shifts the self-intersections into infinity''
and obtains%
\footnote{In the proof of Freedman \cite{Fre:82}, the main complications come
from the lack of control about this process. %
} the standard open 2-handle $(D_{\infty}^{2}\times\mathbb{R}^{2},\partial D_{\infty}^{2}\times\mathbb{R}^{2})$.

A Casson handle is specified up to (orientation preserving) diffeomorphism
(of pairs) by a labeled finitely-branching tree with base-point {*},
having all edge paths infinitely extendable away from {*}. Each edge
should be given a label $+$ or $-$ whereas each vertex corresponds
to a kinky handle. The self-plumbing number of that kinky handle equals
the number of branches leaving the vertex. The sign on each branch
corresponds to the sign of the associated self plumbing. The whole
process generates a tree with infinite many levels. In principle,
every tree with a finite number of branches per level realizes a corresponding
Casson handle. Each building block of a Casson handle, the ``kinky''
handle with $n$ kinks%
\footnote{The number of end-connected sums is exactly the number of self intersections
of the immersed two handle.%
}, is diffeomorphic to the $n-$times boundary-connected sum $\natural_{n}(S^{1}\times D^{3})$
(see appendix \ref{sec:Connected-and-boundary-connected}) with two
attaching regions. Technically speaking, one region is a tubular neighborhood
of band sums of Whitehead links connected with the previous block.
The other region is a disjoint union of the standard open subsets
$S^{1}\times D^{2}$ in $\#_{n}S^{1}\times S^{2}=\partial(\natural_{n}S^{1}\times D^{3})$
(this is connected with the next block).

\subsection{3-manifolds and exotic $\mathbb{R}^{4}$}

We described the construction of large and small exotic $\mathbb{R}^{4}$'s
above. Apart from the different constructions, there are some common
properties of exotic $\mathbb{R}^{4}$'s which will be discussed now.
Usually there are arbitrary splittings of the $\mathbb{R}^{4}$ into
$\mathbb{R}^{3}\times\mathbb{R}$ or $\mathbb{R}\times\mathbb{R}\times\mathbb{R}\times\mathbb{R}$.
But every manifold of dimension smaller than 4 has a unique smoothness
structure. Therefore $\mathbb{R}^{3}\times\left\{ t\right\} $ has
a unique smoothness structure for every $t\in\mathbb{R}$ and also
the whole $\mathbb{R}^{3}\times\mathbb{R}$. A standard smoothness
structure on $\mathbb{R}^{4}$ is uniquely characterized by the fact
that the smoothness structure respects the decomposition $\mathbb{R}^{3}\times\mathbb{R}^{1}$.
But more is true: also every splitting $\Sigma\times\mathbb{R}$ using
a contractable 3-manifold $\Sigma$ is diffeomorphic to $\mathbf{R}^{4}$
(standard $\mathbb{R}^{4}$), see \cite{Chernov2012}. Expressed in
physical terms: there is no globally hyperbolic metric on any exotic
$\mathbb{R}^{4}$. But by using foliation theory, every exotic $\mathbb{R}^{4}$
admits a codimension-1 foliation (or there is a non-vanishing vector
field which can be used to define a Lorentz metric, see \cite{Ste:99}).
Therefore there must be a connection between exotic $\mathbb{R}^{4}$
and (non-trivial) codimension-1 foliations.

Another characterizing property of all known exotic $\mathbb{R}^{4}$
is the existence of a compact subset $Q\subset R^{4}$ which cannot
be surrounded by a smoothly embedded 3-sphere. To express it differently,
the exotic $\mathbb{R}^{4}$ does not contain 3-spheres surrounding
$Q$. This fact will be the central point in our argumentation because
it allows to consider non-trivial 3-manifolds (i.e. not homeomorphic
to the 3-sphere). For us, it is enough to obtain a representative
(i.e. a 3-manifold) for an exotic $\mathbb{R}^{4}$. We are not interested
in the construction of a unique 3-manifold (characterizing the smoothness
structure). Instead we will show that exotic smoothness is the source
of non-trivial 3-manifolds $\Sigma$ which are embedded. In contrast,
for any compact subset $Q$ of the standard $\mathbb{R}^{4}$ there
is always a smoothly embedded 3-sphere surrounding $Q$. Of course
the 3-manifold $\Sigma$ can be also embedded in the standard $\mathbb{R}^{4}$
(at least in case of the small exotic $\mathbb{R}^{4}$) but we will
state that the exotic $\mathbb{R}^{4}$ does not contain 3-spheres
surrounding $Q$ (for all known examples of exotic $\mathbb{R}^{4}$).
Furthermore, the 3-manifold $\Sigma$ surrounding the compact subset
$Q$ in the exotic $\mathbb{R}^{4}$ separates $Q$ from infinity.
\emph{In the following we will denote the 3-manifold surrounding $Q$
by $\Sigma$.} Much of the following material can be found in the
section 9.4 of the book \cite{GomSti:97} as well in the paper \cite{Ganzel2006}.

\subsubsection{Large Exotic $\mathbb{R}^{4}$}

There are only implicit examples of large exotic $\mathbb{R}^{4}$'s
and our construction (\ref{eq:decomposition-large-exotic-R4}) is
in principle not very useful. Let $R^{4}$ be a large exotic $\mathbb{R}^{4}$
and $Q\subset R^{4}$ a compact subset of codimension $0$. In \cite{Ganzel2006},
a large exotic $\mathbb{R}^{4}$ called $R_{1}$ was constructed (using
the co-called K3-surface) so that any possible $Q$ is surrounded
by a 3-manifold $\Sigma$ with first Betti number $b_{1}$ at least
$b_{1}=3$ in contrast to the 3-sphere with $b_{1}=0$.

\subsubsection{Small Exotic $\mathbb{R}^{4}$}

\begin{figure}
\begin{center}\includegraphics[scale=0.2]{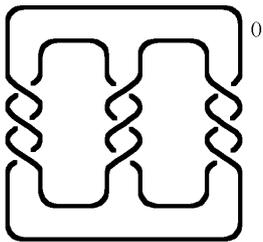}\end{center}

\caption{$(-3,3,-3)$ pretzel knot or knot $9_{46}$ (Rolfson notation)\label{fig:-pretzel-knot-3-3-3}}

\end{figure}

In case of a small exotic $\mathbb{R}^{4}$, there are some explicit
constructions, see \cite{BizGom:96}. Therefore, it is not surprising
that there is an explicit construction (or one possible representative)
for the 3-manifold $\Sigma$ surrounding a possible compact subset
$Q$. It is a 3-manifold with $b_{1}=1$ ($0-$framed surgery along
the $(-3,3,-3)$ pretzel knot, see Fig. \ref{fig:-pretzel-knot-3-3-3},
in the simplest case or the $n-$fold untwisted Whitehead double of
it). We will fix this 3-manifold $\Sigma$ in the following. 

\subsection{The Einstein-Hilbert action}

In this section we will discuss the (Euclidean) Einstein-Hilbert action
functional 
\begin{equation}
S_{EH}(M)=\intop_{M}R\sqrt{g}\: d^{4}x\label{eq:EH-action}
\end{equation}
of the 4-manifold $M$ and fix the Ricci-flat metric $g$ as solution
of the vacuum field equations of the exotic 4-manifold. If $M$ is
an exotic $\mathbb{R}^{4}$ we have to comment about the existence
of a metric $g$. At first, there is a result of Taylor \cite{Taylor2005}
that all known exotic $\mathbb{R}^{4}$'s admit no proper Lipschitz
functions having bounded critical values. This property is very restrictive
but does not forbid to introduce a smooth metric $g$ (at least locally).
Furthermore, it confirmed implicitly a result of us \cite{AsselmeyerKrol2014}
that a cosmological model based on an exotic $\mathbb{R}^{4}$ (or
better on its exotic end $S^{3}\times\mathbb{R}$) showed an inflationary
behavior, i.e. with no bounded curvature etc. 

In the following we will argue to obtain an additional contribution
to the action functional coming from exotic smoothness. This contribution
uses the property of all known exotic $\mathbb{R}^{4}$ (see the previous
subsection above), i.e. the existence of a non-trivial compact 3-manifold
separating a compact submanifold from infinity. Now consider a compact
4-dimensional submanifold $Q\subset R^{4}$ of an exotic $\mathbb{R}^{4}$
(large or small). Then $Q$ is surrounded by a 3-manifold $\Sigma$
which is not diffeomorphic to the 3-sphere $S^{3}$.This 3-manifold
$\Sigma$ divides the exotic $\mathbb{R}^{4}$ into two parts
\begin{equation}
R^{4}=\overline{U}(Q)\cup_{\partial\overline{U}(Q)=\Sigma}\left(R^{4}\setminus\overline{U}(Q)\right)\label{eq:relation-exotic}
\end{equation}
where $\overline{U}(Q)$ is a closed neighborhood of $Q$ with boundary
$\Sigma=\partial\overline{U}(Q)$. For the parts of the decomposition
we obtain the action functionals 
\begin{eqnarray*}
S_{EH}(R^{4}\setminus\overline{U}(Q)) & = & \intop_{R^{4}\setminus\overline{U}(Q)}R\sqrt{g}\, d^{4}x+\intop_{\partial\overline{U}(Q)}H\sqrt{h}d^{3}x\\
S_{EH}(\overline{U}(Q)) & = & \intop_{\overline{U}(Q)}R\sqrt{g}d^{4}x-\intop_{\partial\overline{U}(Q)}H\sqrt{h}d^{3}x
\end{eqnarray*}
including the contribution of the boundary $\partial\overline{U}(Q)=\Sigma$
with respect to different orientations and $H$ is the trace of the
second fundamental form (mean curvature) of the boundary in the metric
$g$ (see \cite{Ashtekar08,Ashtekar08a} for the discussion of the
boundary terms). Interestingly, this decomposition is independent
of the class, large or small, of the exotic $\mathbb{R}^{4}$. In
the following we will discuss the boundary term, i.e. we can reduce
the problem to the discussion of the action 
\begin{equation}
S_{EH}(\Sigma)=\intop_{\Sigma}H\,\sqrt{h}\, d^{3}x\label{eq:action fermi}
\end{equation}
along the boundary $\Sigma$ (a 3-manifold). It is a surprise that
this integral agrees with the Dirac action of a spinor describing
the (embedded) boundary, see below.

\section{Dirac action and 3-manifolds\label{sec:Dirac-action-3MF}}

In the following we will show that the action (\ref{eq:action fermi})
over a 3-manifold $\Sigma$ is equivalent to the the Dirac action
of a spinor over $\Sigma$. At first we will consider the general
case of an embedding of a 3-manifold into a 4-manifold. Let $\iota:\Sigma\hookrightarrow M$
be an embedding of the 3-manifold $\Sigma$ into the 4-manifold $M$
with the normal vector $\vec{N}$. A small neighborhood $U_{\epsilon}$
of $\iota(\Sigma)\subset M$ looks like $U_{\epsilon}=\iota(\Sigma)\times[0,\epsilon]$.
Furthermore we identify $\Sigma$ and $\iota(\Sigma)$ ($\iota$ is
an embedding). Every 3-manifold admits a spin structure with a {\em spin bundle}, 
i.e. a principal $Spin(3)=SU(2)$ bundle (spin bundle) as
a lift of the frame bundle (principal $SO(3)$ bundle associated to
the tangent bundle). There is a (complex) vector bundle associated
to the spin bundle (by a representation of the spin group), called
{\em spinor bundle} $S_{\Sigma}$. A section in the spinor bundle
is called a spinor field (or a spinor). In case of a 4-manifold, we
have to assume the existence of a spin structure. But for a manifold
like $\mathbb{R}^{4}$, there is no restriction, i.e. there is always
a spin structure and a spinor bundle $S_{M}$. In general, the unitary
representation of the spin group in $D$ dimensions is $2^{[D/2]}$-dimensional.
From the representational point of view, a spinor in 4 dimensions
is a pair of spinors in dimension 3. Therefore, the spinor bundle
$S_{M}$ of the 4-manifold splits into two sub-bundles $S_{M}^{\pm}$
where one subbundle, say $S_{M}^{+},$ can be related to the spinor
bundle $S_{\Sigma}$ of the 3-manifold. Then the spinor bundles are
related by $S_{\Sigma}=\iota^{*}S_{M}^{+}$ with the same relation
$\phi=\iota_{*}\Phi$ for the spinors ($\phi\in\Gamma(S_{\Sigma})$
and $\Phi\in\Gamma(S_{M}^{+})$). Let $\nabla_{X}^{M},\nabla_{X}^{\Sigma}$
be the covariant derivatives in the spinor bundles along a vector
field $X$ as section of the bundle $T\Sigma$. Then we have the formula
\begin{equation}
\nabla_{X}^{M}(\Phi)=\nabla_{X}^{\Sigma}\phi-\frac{1}{2}(\nabla_{X}\vec{N})\cdot\vec{N}\cdot\phi\label{eq:covariant-derivative-immersion}
\end{equation}
with the obvious embedding $\phi\mapsto\left(\begin{array}{c}
\phi\\
0
\end{array}\right)=\Phi$ of the spinor spaces from the relation $\phi=\iota_{*}\Phi$. The
expression $\nabla_{X}\vec{N}$ is the second fundamental form of
the embedding where the trace $tr(\nabla_{X}\vec{N})=2H$ is related
to the mean curvature $H$. Then from (\ref{eq:covariant-derivative-immersion})
one obtains the following relation between the corresponding Dirac
operators 
\begin{equation}
D^{M}\Phi=D^{\Sigma}\phi-H\phi\label{eq:relation-Dirac-3D-4D}
\end{equation}
with the Dirac operator $D^{\Sigma}$ on the 3-manifold $\Sigma$.
This relation (as well as (\ref{eq:covariant-derivative-immersion}))
is only true for the small neighborhood $U_{\epsilon}$ where the
normal vector points is parallel to the vector defined by the coordinates
of the interval $[0,\epsilon]$ in $U_{\epsilon}$. In \cite{AsselmeyerRose2012},
we extend the spinor representation of an immersed surface into the
3-space to the immersion of a 3-manifold into a 4-manifold according
to the work in \cite{Friedrich1998}. Then the spinor $\phi$ defines
directly the embedding (via an integral representation) of the 3-manifold.
Then the restricted spinor $\Phi|_{\Sigma}=\phi$ is parallel transported
along the normal vector and $\Phi$ is constant along the normal direction
(reflecting the product structure of $U_{\epsilon}$). But then the
spinor $\Phi$ has to fulfill 
\begin{equation}
D^{M}\Phi=0\label{eq:Dirac-equation-4D}
\end{equation}
in $U_{\epsilon}$ i.e. $\Phi$ is a parallel spinor. Finally we get
\begin{equation}
D^{\Sigma}\phi=H\phi\label{eq:Dirac3D-mean-curvature}
\end{equation}
with the extra condition $|\phi|^{2}=const.$ (see \cite{Friedrich1998}
for the explicit construction of the spinor with $|\phi|^{2}=const.$
from the restriction of $\Phi$). Then we can express the action (\ref{eq:action fermi})
by using (\ref{eq:Dirac3D-mean-curvature}) to obtain
\begin{equation}
\intop_{\Sigma}H\,\sqrt{h}\, d^{3}x=\intop_{\Sigma}\bar{\phi}\, D^{\Sigma}\phi\,\sqrt{h}d^{3}x\label{eq:relation-mean-curvature-action-to-dirac-action}
\end{equation}
using $|\phi|^{2}=const.$

\subsection{Deformation of the Embedding as seen by the Dirac operator}

Now we will discuss the deformation of an embedding using a diffeomorphism.
Let $I:\Sigma\hookrightarrow M$ be an embedding of $\Sigma$ (3-manifold)
into $M$ (4-manifold). A deformation of an embedding $I':\Sigma'\hookrightarrow M'$
are diffeomorphisms $f:M\to M'$ and $g:\Sigma\to\Sigma'$ of $M$
and $\Sigma$, respectively, so that 
\[
I\circ f=g\circ I'\,.
\]
One of the diffeomorphism (say $f$) can be absorbed into the definition
of the embedding and we are left with one diffeomorphism $g\in Diff(\Sigma)$
to define the deformation of the embedding $I$. But as stated above,
the embedding is directly related to the spinor $\phi$ on $\Sigma$
fulfilling the Dirac equation. Therefore we have to discuss the action
of the diffeomorphism group $Diff(\Sigma)$ on the Hilbert space of
$L^{2}-$spinors fulfilling the Dirac equation. This case was considered
in the literature \cite{SpinorsDiffeom2013}. The spinor space $S_{g,\sigma}(\Sigma)$
on $\Sigma$ depends on two ingredients: a (Riemannian) metric $g$
and a spin structure $\sigma$ (labeled by the number of elements
in $H^{1}(\Sigma,\mathbb{Z}_{2})$). Let us consider the group of
orientation-preserving diffeomorphism $Diff^{+}(\Sigma)$ acting on
$g$ (by pullback $f^{*}g$) and on $\sigma$ (by a suitable defined
pullback $f^{*}\sigma$). The Hilbert space of $L^{2}-$spinors of
$S_{g,\sigma}(\Sigma)$ is denoted by $H_{g,\sigma}$. Then according
to \cite{SpinorsDiffeom2013}, any $f\in Diff^{+}(\Sigma)$ leads
in exactly two ways to a unitary operator $U$ from $H_{g,\sigma}$
to $H_{f^{*}g,f^{*}\sigma}$. The (canonically) defined Dirac operator
is equivariant with respect to the action of $U$ and the spectrum
is invariant under (orientation-preserving) diffeomorphisms. But by
the discussion above, we also do not change the embedding by a diffeomorphism.
So, our whole approach is independent of a particular coordinate system
i.e. $\Phi$ is a parallel spinor. In \cite{AsselmeyerRose2012},
we used an one-parameter family of surface embeddings instead of a
3-manifold embedding. But the result remains the same: the 4-dimensional
spinor $\Phi$ is a parallel spinor.

\subsection{The extension to the 4-dimensional Dirac action}

Above we obtained a relation (\ref{eq:relation-Dirac-3D-4D}) between
a 3-dimensional spinor $\phi$ on the 3-manifold $\Sigma$ fulfilling
a Dirac equation $D^{\Sigma}\phi=H\phi$ (determined by the embedding
$\Sigma\to M$ into a 4-manifold $M$) and a 4-dimensional spinor
$\Phi$ on a 4-manifold $M$ with fixed chirality ($\in\Gamma(S_{M}^{+})$
or $\in\Gamma(S_{M}^{-})$) fulfilling the Dirac equation $D^{M}\Phi=0$.
At first we consider the variation 
\begin{equation}
\delta\intop_{\Sigma}\bar{\phi}D^{\Sigma}\phi\:\sqrt{g}\, d^{3}x=0\label{eq:3D-variation}
\end{equation}
of the 3-dimensional action leading to the Dirac equations 
\begin{equation}
D^{\Sigma}\phi=0\quad D^{\Sigma}\bar{\phi}=0\label{eq:3D-Dirac-equation}
\end{equation}
or to 
\[
H=0\,,
\]
a characterization of the embedding $\Sigma\to M$ with minimal mean
curvature. This variation can be understood as a variation of the
embedding. In contrast, the extension of the spinor $\phi$ (as solution
of (\ref{eq:3D-Dirac-equation})) to the 4-dimensional spinor $\Phi$
by using the embedding 
\begin{equation}
\Phi=\left(\begin{array}{c}
\phi\\
0
\end{array}\right)\label{eq:embedding-spinor-3D-4D}
\end{equation}
can be only seen as embedding, if (and only if) the 4-dimensional
Dirac equation 
\begin{equation}
D^{M}\Phi=0\label{eq:4D-Dirac-equation}
\end{equation}
on $M$ is fulfilled (using relation (\ref{eq:relation-Dirac-3D-4D})).
This Dirac equation is obtained by varying the action 
\begin{equation}
\delta\intop_{M}\bar{\Phi}D^{M}\Phi\sqrt{g}\: d^{4}x=0\label{eq:4D-variation}
\end{equation}
Importantly, this variation has a different interpretation in contrast
to varying the 3-dimensional action. Both variations look very similar.
But in (\ref{eq:4D-variation}) we vary over smooth maps $\Sigma\to M$
which are not embeddings (i.e. represented by spinors $\Phi$ with
$D^{M}\Phi\not=0$). Only the choice of the extremal action selects
the embedding among other smooth maps. In particular the spinor $\Phi$
(as solution of the 4-dimensional Dirac equation) is localized at
the embedded 3-manifold $\Sigma$ (with respect to the embedding (\ref{eq:embedding-spinor-3D-4D})).
The 3-manifold $\Sigma$ moves along the normal vector (see the relation
(\ref{eq:covariant-derivative-immersion}) between the covariant derivatives
representing a parallel transport).

\subsection{Fermions as knot complements\label{sub:Fermions-as-knot-complement}}

In the previous subsections we presented a formalism to describe the
embedding of a 3-manifold into a 4-manifold by using a spinor. But
one may ask whether the spinor is really connected with a field of
spin $\frac{1}{2}$. To answer this question we have to analyze the
structure of 3-manifolds. In short, every 3-manifold is the sum of
prime 3-manifolds where a subclass (irreducible 3-manifolds) splits
into hyperbolic and graph manifolds.

\begin{figure}
\includegraphics{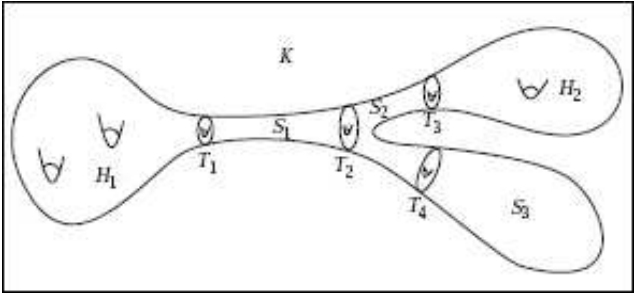}

\caption{Torus (JSJ-) decomposition, $H_{i}$ hyperbolic manifold, $S_{i}$
Graph-manifold, $T_{i}$ Tori \label{fig:Torus-decomposition}}
\end{figure}

A connected 3-manifold $\Sigma$ is prime if it cannot be obtained
as a connected sum of two manifolds $\Sigma_{1}\#\Sigma_{2}$ (see
the appendix \ref{sec:Connected-and-boundary-connected} for the definition)
neither of which is the 3-sphere $S^{3}$ (or, equivalently, neither
of which is the homeomorphic to $\Sigma$). Examples are the 3-torus
$T^{3}$ and $S^{1}\times S^{2}$ but also the Poincare sphere. According
to \cite{Mil:62}, any compact, oriented 3-manifold is the connected
sum of an unique (up to homeomorphism) collection of prime 3-manifolds
(prime decomposition). A subset of prime manifolds are the irreducible
3-manifolds. A connected 3-manifold is irreducible if every differentiable
submanifold $S$ homeomorphic to a sphere $S^{2}$ bounds a subset
$D$ (i.e. $\partial D=S$) which is homeomorphic to the closed ball
$D^{3}$. The only prime but reducible 3-manifold is $S^{1}\times S^{2}$.
For the geometric properties (to meet Thurstons geometrization theorem)
we need a finer decomposition induced by incompressible tori. A properly
embedded connected surface $S\subset\Sigma$ is called 2-sided%
\footnote{The \textquoteleft{}sides\textquoteright{} of $S$ then correspond
to the components of the complement of $S$ in a tubular neighborhood
$S\times[0,1]\subset N$.%
} if its normal bundle is trivial, and 1-sided if its normal bundle
is nontrivial. A 2-sided connected surface $S$ other than $S^{2}$
or $D^{2}$ is called incompressible if for each disk $D\subset\Sigma$
with $D\cap S=\partial D$ there is a disk $D'\subset S$ with $\partial D'=\partial D$.
The boundary of a 3-manifold is an incompressible surface. Most importantly,
the 3-sphere $S^{3}$, $S^{2}\times S^{1}$ and the 3-manifolds $S^{3}/\Gamma$
with $\Gamma\subset SO(4)$ a finite subgroup do not contain incompressible
surfaces. The class of 3-manifolds $S^{3}/\Gamma$ (the spherical
3-manifolds) includes cases like the Poincare sphere ($\Gamma=I^{*}$
the binary icosaeder group) or lens spaces ($\Gamma=\mathbb{Z}_{p}$
the cyclic group). Let $K_{i}$ be irreducible 3-manifolds containing
incompressible surfaces then we can $N$ split into pieces (along
embedded $S^{2}$)
\begin{equation}
\Sigma=K_{1}\#\cdots\#K_{n_{1}}\#_{n_{2}}S^{1}\times S^{2}\#_{n_{3}}S^{3}/\Gamma\,,\label{eq:prime-decomposition}
\end{equation}
where $\#_{n}$ denotes the $n$-fold connected sum and $\Gamma\subset SO(4)$
is a finite subgroup. The decomposition of $N$ is unique up to the
order of the factors. The irreducible 3-manifolds $K_{1},\ldots,\, K_{n_{1}}$
are able to contain incompressible tori and one can split $K_{i}$
along the tori into simpler pieces $K=H\cup_{T^{2}}G$ \cite{JacSha:79}
(called the JSJ decomposition). The two classes $G$ and $H$ are
the graph manifold $G$ and hyperbolic 3-manifold $H$ (see Fig. \ref{fig:Torus-decomposition}).
The hyperbolic 3-manifold $H$ has a torus boundary $T^{2}=\partial H$,
i.e. $H$ admits a hyperbolic structure in the interior only. One
property of hyperbolic 3-manifolds is central: Mostow rigidity. As
shown by Mostow \cite{Mos:68}, every hyperbolic $n-$manifold $n>2$
with finite volume has this property: \emph{Every diffeomorphism (especially
every conformal transformation) of a hyperbolic $n-$manifold with
finite volume is induced by an isometry.} Therefore one cannot scale
a hyperbolic 3-manifold and the volume is a topological invariant.
Together with the prime and JSJ decomposition
\begin{equation}
\Sigma=\left(H_{1}\cup_{T^{2}}G_{1}\right)\#\cdots\#\left(H_{n_{1}}\cup_{T^{2}}G_{n_{1}}\right)\#_{n_{2}}S^{1}\times S^{2}\#_{n_{3}}S^{3}/\Gamma\,,\label{eq:general-decomposition-3MF}
\end{equation}
we can discuss the geometric properties central to Thurstons geometrization
theorem (as proved by Perelman): \emph{Every oriented closed irreducible
3-manifold can be cut along tori (JSJ decomposition), so that the
interior of each of the resulting manifolds has a geometric structure
with finite volume.} Now, we have to introduce the term 'geometric
structure'. A model geometry is a simply connected smooth manifold
$X$ together with a transitive action of a Lie group $G$ on $X$
with compact stabilizers. A geometric structure on a manifold $\Sigma$
is a diffeomorphism from $\Sigma$ to $X/{\Gamma}$ 
for some model geometry $X$, where $\Gamma$
is a discrete subgroup of $G$ acting freely on $X$. It is a surprising
fact that there are also a finite number of three-dimensional model
geometries, i.e. 8 geometries with the following models: spherical
$(S^{3},O_{4}(\mathbb{R}))$, Euclidean $(\mathbb{E}^{3},O_{3}(\mathbb{R})\ltimes\mathbb{R}^{3})$,
hyperbolic $(\mathbb{H}^{3},O_{1,3}(\mathbb{R})^{+})$, mixed spherical-Euclidean
$(S^{2}\times\mathbb{R},O_{3}(\mathbb{R})\times\mathbb{R}\times\mathbb{Z}_{2})$,
mixed hyperbolic-Euclidean $(\mathbb{H}^{2}\times\mathbb{R},O_{1,3}(\mathbb{R})^{+}\times\mathbb{R}\times\mathbb{Z}_{2})$
and 3 exceptional cases called $\tilde{SL}_{2}$ (twisted version
of $\mathbb{H}^{2}\times\mathbb{R}$), NIL (geometry of the Heisenberg
group as twisted version of $\mathbb{E}^{3}$), SOL (split extension
of $\mathbb{R}^{2}$ by $\mathbb{R}$, i.e. the Lie algebra of the
group of isometries of the 2-dimensional Minkowski space). We refer
to \cite{Scott1983} for the details.

There are three main parts in the possible decomposition of a 3-manifold:
$S^{1}\times S^{2}$, $S^{3}/\Gamma$ and $H_{i}\cup_{T^{2}}G$. We
will introduce the physical property spin in the next section. In
this paper we are interested in fermions with spin $\frac{1}{2}$.
Using the definition of spin $\frac{1}{2}$ in the next section, the
two manifolds $S^{1}\times S^{2}$ and $S^{3}/\Gamma$ have a spin
different from $\frac{1}{2}$. Therefore for the moment, the two manifolds
$S^{1}\times S^{2}$ and $S^{3}/\Gamma$ can be ruled out. We will
discuss the case of higher spins in the next section. Then we are
left with $H_{i}\cup_{T^{2}}G_{i}$. In case of a small exotic $\mathbb{R}^{4}$,
we have an explicit expression of the 3-manifold $\Sigma$: it is
the $0-$framed surgery along $n$th untwisted Whitehead double of
the pretzel knot $(-3,3-3)$ (also known as $9_{46}$ knot in Rolfson
notation, see \cite{Rol:76}). To see it explicitly, we fix $n=1$
and consider the $0-$framed surgery along the untwisted Whitehead
double $K_{Wh,9_{46}}$ of the $(-3,3-3)$ pretzel knot to get a 3-manifold
$\Sigma$. By definition, $\Sigma$ is decomposed like
\begin{equation}
\Sigma=\left(S^{3}\setminus(K_{Wh,9_{46}}\times D^{2})\right)\cup_{T^{2}}(D^{2}\times S^{1})\label{eq:decomposition-3MF}
\end{equation}
with the identity map as gluing map ($D^{2}\times S^{1}$ is a graph
manifold). The 3-manifold $S^{3}\setminus(K_{Wh,9_{46}}\times D^{2})$
is the complement of the Whitehead double $K_{Wh,9_{46}}$. According
to \cite{Budney2006}, one can decompose this knot complement according
to the JSJ-decomposition into
\begin{equation}
S^{3}\setminus(K_{Wh,9_{46}}\times D^{2})=\left(S^{3}\setminus(Wh\times D^{2})\right)\cup_{T^{2}}\left(S^{3}\setminus(K_{9_{46}}\times D^{2})\right)\label{eq:JSJ-decomposition-knot-complement}
\end{equation}
i.e. in the complement of the Whitehead link $Wh$ connected to the
complement of the pretzel knot $K_{9_{46}}$. It is known that both
complements are hyperbolic 3-manifolds. Finally we see that $\Sigma$
can be decomposed into $\Sigma=H_{1}\cup_{T^{2}}H_{2}\cup_{T^{2}}G$
with $H_{1}=S^{3}\setminus(K_{9_{46}}\times D^{2})$, $H_{2}=S^{3}\setminus(Wh\times D^{2})$
and $G=D^{2}\times S^{1}$. From the physical point of view it is
natural to identify the simplest (irreducible) parts of the 3-manifold
with the constitutes of matter. Finally we conjecture by using this
example:\\
\emph{Conjecture: The constitutes of matter are represented by complements
$S^{3}\setminus(D^{2}\times K)$ of knots $K$ with dynamics determined
by the Dirac equation (\ref{eq:4D-Dirac-equation}).}\\
In the next section we will support this conjecture by using physical
arguments. But currently this assumption is not a large restriction.
There are infinitely many knots and we do not know which knot represents
the electron or neutrino. But for knot complements, there is a simple
division into two classes: knot complements admitting a homogenous,
hyperbolic metric (a metric of constant negative curvature in every
direction) and knot complements not admitting such a metric. In \cite{AsselmeyerRose2012},
we discussed the non-hyperbolic case and showed that the corresponding
3-manifolds are representing the interaction. Therefore we are left
with knot complements admitting a hyperbolic metric, called \emph{hyperbolic
knot complements}. In the next section we will show that these knot
complements have the right properties to describe fermions.

\section{The physical interpretation\label{sec:The-physical-interpretation}}

In this section we will discuss the physical interpretation of the
mathematical results above including the limits of this approach.
In particular we will prove the conjecture that hyperbolic knot complements,
i.e. 3-manifolds $S^{3}\setminus\left(D^{2}\times K\right)$ admitting
a homogenous, hyperbolic metric, representing the fermions. We used
the spinor representation to express the embedding of the 3-manifold.
Here we will further clarify the following questions: Does the submanifold
(the knot complement) has the properties of a spinor fulfilling the
Dirac equation? Has it also the properties of matter like non-contractability
(state equation $p=0$)? From the physics point of view, we have to
check that the submanifold (=knot complement) has 
\begin{enumerate}
\item spin $\frac{1}{2}$ (with an appropriated definition), 
\item the Dirac equation of motion and 
\item the state equation $p=0$ (non-contractable matter) in the cosmological
context. 
\end{enumerate}
\textbf{ad 1.} We start with the spin. Our definition is inspired
by the work of Friedman and Sorkin \cite{FriedmanSorkin1980}, for
the details we refer to the Appendix \ref{sec:Spin-from-space}. Now
we will looking for a rotation $R(\theta)$ (rotation w.r.t. an angle
$\theta$) which acts on the 4-dimensional spinor $\Phi$. Because
of the embedding (\ref{eq:embedding-spinor-3D-4D}), it is enough
to consider the action on the 3-dimensional spinor $\phi$. Then a
rotation as element of $SO(3)$ must be represented by a diffeomorphism,
i.e. we have the representation $R:SO(3)\to Diff(\Sigma)$ where $R(\theta)$
is a one-parameter subgroup of diffeomorphisms. We call $\phi$ a
spinor if 
\[
\phi\circ R(2\pi)^{*}=-\phi\qquad\mbox{or}\qquad R(2\pi)=-1
\]
in the notation of Appendix \ref{sec:Spin-from-space}. From the topological
point of view, this rotation is located in the component of the diffeomorphism
group which is not connected to the identity. The existence of these
rotations is connected to the complexity of the 3-manifold. As shown
by Hendriks \cite{Hendriks1977}, these rotations do \textbf{not}
exist in sums of 3-manifolds containing 
\begin{itemize}
\item $\mathbb{R}P^{2}\times S^{1}$ with the 2-dimensional real projective
space $\mathbb{R}P^{2}$ 
\item $S^{2}$ fiber bundle over $S^{1}$ and 
\item for 3-manifolds with finite fundamental group having a cyclic 2-Sylow
subgroup%
\footnote{A 2-Sylow subgroup of a finite group (here the fundamental group)
is a subgroup whose order is a power of $2$ (possibly $2^{0}$) and
which is properly contained in no larger Sylow subgroup. We note that
all 2-Sylow subgroups of a given group are isomorphic.%
}. 
\end{itemize}
In case of hyperbolic 3-manifolds (the knot complements) one has an
infinite fundamental group and therefore it has spin $\frac{1}{2}$.
What about higher spins? A realistic approach to higher spins is the
consideration of symmetries w.r.t. the rotation group. Again we use
the action on the 3-dimensional spinor. Then spin $1$ is the action
$\phi\circ R(2\pi)^{*}=\phi$ and spin $2$ is given by $\phi\circ R(\pi)^{*}=\phi$
etc. But in contrast to the spin $\frac{1}{2}$ case, these conditions
are not a large restriction. Examples are the (non-trivial) torus
bundles for spin $1$ and $S^{2}\times[0,1]$ for spin $2$ (ignoring
the orientation). \\
\textbf{ad 2.} This part was already shown. Using the variation (\ref{eq:4D-variation})
we obtain the 4-dimensional Dirac equation (\ref{eq:4D-Dirac-equation})
in case of an embedding. Then the spinor is directly interpretable
as the embedding, see above. \\
 \textbf{ad 3.} In cosmology, one has to introduce a state equation
\[
p=w\cdot\rho
\]
between the pressure and the energy density. Matter as formed by fermions
is characterized by the state equation $p=0$ or $w=0$. Equivalently,
matter is incompressible and the energy density $\rho\sim a^{-3}$
scales like the inverse volume of the 3-space w.r.t. scaling factor
$a$. By the hypothesis above, we consider the complement of the hyperbolic
knot which is a hyperbolic 3-manifold $H$ having a torus boundary
$T^{2}=\partial H$, i.e. $H$ admits a hyperbolic structure in the
interior only. To identify $H$ with matter, it should also have the
property of incompressibility. But what does it mean? As a model of
a cosmos we consider a closed 3-manifold $\Sigma$ (no boundary).
By the decomposition (\ref{eq:general-decomposition-3MF}) we have
to glue the hyperbolic manifold $H$ \emph{at least} to one graph
manifold $G$ to get a closed 3-manifold $\Sigma$. Adding more components
like $S^{1}\times S^{2}$ or $S^{3}/\Gamma$ is always possible. But
the\emph{ minimal model for a cosmos} $\Sigma$ is given by 
\begin{equation}
\Sigma=H\cup_{T^{2}}G\label{eq:minimal-model-cosmos}
\end{equation}
where the two manifolds $H$ and $G$ have a common boundary, the
torus. $H$ represents the matter (by our assumption) and $G$ is
the surrounding space. Furthermore we assume that $\Sigma$ scales
w.r.t. the scaling factor $a$, i.e. $vol(\Sigma)\sim a^{3}$. The
energy density is the total energy $E_{\Sigma}$ of the matter per
volume or 
\[
\rho=\frac{E_{\Sigma}}{vol(\Sigma)}\quad.
\]
The total energy $E_{\Sigma}$ is related to the scalar curvature,
see appendix \ref{sec:Scalar-curvature-and-energy-density}. Using
(\ref{eq:total-energy-constant-curvature}), we obtain for the total
energy $E_{H}$ of the hyperbolic 3-manifold $H$ 
\[
E_{H}=vol(H)\cdot\left(\frac{1}{\kappa}R_{H}+\rho_{c}\right)\quad.
\]
Therefore we will get the scaling law $\rho_{H}=E_{H}/vol(\Sigma)\sim a^{-3}$
only for $E_{H}\sim a^{0}$ by using the assumption $vol(\Sigma)\sim a^{3}$.
It is interesting that the properties of hyperbolic 3-manifolds agree
with this demand. Because of Mostow rigidity, one cannot scale a finite-volume,
hyperbolic 3-manifold. Then the volume $vol(H)$ and the curvature
of $H$ are topological invariants. But then $E_{H}$ is also a topological
invariant with the scaling behavior $E_{H}\sim a^{0}$ of a topological
invariant (see appendix \ref{sec:Scalar-curvature-and-energy-density}).
Therefore the scaling of the volume $vol(\Sigma)\sim a^{3}$ is caused
by the manifold $G$ (surrounding space). Finally we obtain the scaling
of matter in cosmology to be $a^{-3}$ or $w=0$ confirming that matter
can be geometrically represented by hyperbolic 3-manifolds. But at
this stage of the work, one may ask whether the knot complements determine
the matter content of the universe (at least in principle). The 4-spinor
$\Phi$ fulfilling the Dirac equation (\ref{eq:Dirac-equation-4D})
describes the embedding of the 3-manifold (as well the knot complement).
But $\Phi$ is a fermion field in physics view which does not contain
specific properties of the corresponding fermion like charge or mass.
By using the equation (\ref{eq:Dirac3D-mean-curvature}), we are able
to interpret the mean curvature as 3-momentum (up to Plancks constant).
Above we showed that the hyperbolic knot complement has the properties
of a fermion. For a realistic model of matter in the cosmos, we need
a lot of fermions. To produce them, we have to choose a more complex
(small) exotic $\mathbb{R}^{4}$ containing a Casson handle with many
branches and a larger neighborhood to obtain a more complex 3-manifold
$\Sigma$. As an example consider the Casson handle produced from
the dual tree (every vertex has a branching into two vertices) where
every edge has the same label ($+$ or $-$). Furthermore we consider
the $n$th Whitehead double%
\footnote{To express the branching, one needs a more complex Whitehead link
containing more circles. For the details consult the book \cite{GomSti:97}.%
}. Then the decomposition (\ref{eq:decomposition-3MF}) of the 3-manifold
does not change but contains now a knot complement of a more complex
knot. Then the number of knot complements $C(K_{9_{46}})$ of the
knot $9_{46}$ (or the pretzel knot) in the 3-manifold is now $2^{n}$.
Therefore an appropriate choice of $n$ can produce a realistic content
of fermions (as represented by knot complements). Currently we have
no idea which properties of the knot complement are connected to the
properties of the fermion but we think these properties are not so
strongly connected to the particular knot complement.\\
Finally: \emph{Fermions are represented by hyperbolic knot complements.}

\section{The Brans conjecture: generating sources of gravity\label{sec:The-Brans-conjecture}}

We only do direct geometric observations within some local, human-scaled
coordinate patch, including, of course, interpolations of signals
received from sources outside this patch. From this, we usually assume
that space-time has the simplest global smoothness structure. Suppose
it does not, so that space-time is exotically smooth. For example,
suppose we observe only a single mass outside our local region and
it looks like a black hole. Normally, we assume we can extrapolate
data arriving in our standard coordinate patch on earth all the way
back to the vicinity of the black hole. We ask: ''what if the smoothness
structure does not allow this?''

This question is at the core of the Brans conjecture. Exotic space-times
like the exotic $\mathbb{R}^{4}$ have the property that there is
no foliation like $\mathbb{R}^{3}\times\mathbb{R}$ otherwise the
space-time has a standard smoothness structure. But all other foliations
break the strong causality, i.e. there is no unique geodesics going
in the future or past (see the discussion in \cite{AsselmeyerRose2012}).
In this paper we will go a step further and will interpret the deviation
of the smoothness structure from the standard smoothness structure
as sources of gravity. In particular we will use the theory above
to identify the sources as fermions.

For that purpose, let us consider an exotic $\mathbb{R}^{4}$ called
$R^{4}$ and the standard $\mathbb{R}^{4}$ called $\mathbf{R}^{4}$.
Let $Q$ be a common 4-dimensional submanifold $Q\subset R^{4}$ and
$Q\subset\mathbf{R}^{4}$. Then there are two closed neighborhoods
$\overline{U}_{R^{4}}(Q)$ and $\overline{U}_{\mathbf{R}^{4}}(Q)$
in $R^{4}$ and $\mathbf{R}^{4}$, respectively. The boundaries of
the closed neighborhoods are different, i.e. $\partial\overline{U}_{R^{4}}(Q)=\Sigma$
and $\partial\overline{U}_{\mathbf{R}^{4}}(Q)$ can be chosen to be
the 3-sphere $S^{3}$. The action for the closed neighborhoods splits
like

\[
S_{EH}(\overline{U}(Q))=\intop_{\overline{U}(Q)}R\sqrt{g}d^{4}x-\intop_{\partial\overline{U}(Q)}H\sqrt{h}d^{3}x
\]
and we can choose an metric in the interior of the closed neighborhoods
to obtain the same value of the action
\[
\intop_{\overline{U}_{R^{4}}(Q)}R\sqrt{g}d^{4}x=\intop_{\overline{U}_{\mathbf{R}^{4}}(Q)}R\sqrt{g}d^{4}x
\]
for the interior of $\overline{U}_{\mathbf{R}^{4}}(Q)$ and $\overline{U}_{R^{4}}(Q)$.
Then the (formal) difference of the actions will result 
\begin{equation}
S_{EH}(\overline{U}_{R^{4}}(Q))-S_{EH}(\overline{U}_{\mathbf{R}^{4}}(Q))=\intop_{S^{3}}H\sqrt{h}d^{3}x-\intop_{\Sigma}H\sqrt{h}d^{3}x\label{eq:relation-actions}
\end{equation}
 in the difference of the boundary integrals. By the discussion above,
only the action along $\Sigma$ can be physically identified with
the spinor, i.e.
\[
\intop_{\Sigma}H\sqrt{h}\, d^{3}x=\intop_{\Sigma}\overline{\phi}D^{\Sigma}\phi\sqrt{h}\, d^{3}x\:.
\]
Therefore in comparison to $\mathbf{R}^{4}$ (standard $\mathbb{R}^{4}$)
one has an additional term in the action 
\[
S_{EH}(\overline{U}_{R^{4}}(Q))=S_{EH}(\overline{U}_{\mathbf{R}^{4}}(Q))+\intop_{S^{3}}H\sqrt{h}d^{3}x-\intop_{\Sigma}\overline{\phi}D^{\Sigma}\phi\sqrt{h}\, d^{3}x
\]
by using the relation (\ref{eq:relation-actions}) which can be extended
\[
S_{EH}(\overline{U}_{R^{4}}(Q))=S_{EH}(\overline{U}_{\mathbf{R}^{4}}(Q))+\intop_{S^{3}}H\sqrt{h}d^{3}x-\intop_{U_{\epsilon}(\Sigma)}\overline{\Phi}D^{U(\Sigma)}\Phi\sqrt{g}\, d^{4}x
\]
to the 4-dimensional part $U_{\epsilon}(\Sigma)$ where $\Sigma$
embeds. So, what did we showed? It is known that the change of the
smoothness structure results in a change of the geometric properties
(Einstein metrics to non-Einstein metrics \cite{Lebrun96}, for instance).
Above we analyzed this change at the level of action functionals.
Amazingly, there is an additional term which can be written as the
Dirac action. But this term must be the reason for the change of the
geometry, or this term is physically the source of gravity. In case
of the exotic $\mathbb{R}^{4}$, S{\l}adkowski \cite{Sla:99,Sladkowski2001}
obtained a change from the flat $\mathbf{R}^{4}$ (standard) to the
curved $R^{4}$ (exotic). So, exotic smoothness is producing this
representation of fermions including an action of the surrounding
space. 

This result has also some impact on the Brans conjecture, that exotic
smoothness will produce an additional gravitational field. Now we
identified these sources partly as fermion fields. The Brans conjecture
is now concretized: the source of the additional gravitational field
can be fermions. To find all other sources remains a task for the
future.

\section{\label{sec:Outlook}Outlook: how to include fermions in general relativity
by using non-trivial 3-manifolds}

Now we will reverse our argumentation, i.e. we assume a space-time
with spatial component $N$ (space) not containing any matter. By
adding $N\#\Sigma$ a non-trivial 3-manifold $\Sigma$ to the space
$N$, we will obtain an action which contains matter (at least fermions).
Let $M$ be a space-time which will be assumed to look like $M=N\times[0,1)$.
Then the (sourceless) Einstein-Hilbert action of $M$ is given by
\[
S_{EH}(M)=\intop_{M}R\sqrt{g}\: d^{4}x+\intop_{N}H\sqrt{h}\, d^{3}x
\]
where the last term is the boundary term of $M$. In some models one
considers $N\times\mathbb{R}$ and includes the boundary at infinity.
Now we modify the 3-manifold into $N\#\Sigma$ using the model (\ref{eq:minimal-model-cosmos})
and get for the action
\[
\intop_{M'}R\sqrt{g}\: d^{4}x+\intop_{N}H\sqrt{h}\, d^{3}x+\intop_{\Sigma}H\sqrt{h}\, d^{3}x+extra\, terms
\]
for the modified space-time $\tilde{M}=(N\#\Sigma)\times[0,1)$. Here
we neglect the extra terms induced by the definition of the connected
sum $\#$. By the model (\ref{eq:minimal-model-cosmos}), the 3-space
$\Sigma$ contains a hyperbolic knot complement $H$ which is described
by a 3-spinor $\phi$. Then we obtain the action
\[
\intop_{\tilde{M}}R\sqrt{g}\: d^{4}x+\intop_{N}H\sqrt{h}\, d^{3}x+\intop_{\Sigma}\bar{\phi}D^{\Sigma}\phi\sqrt{h}\, d^{3}x
\]
as well as an extension of the 3-spinor $\phi$ to a (chiral) 4-spinor
$\Phi$ in $\tilde{M}$. Finally we get
\[
\intop_{\tilde{M}}R\sqrt{g}\: d^{4}x+\intop_{N}H\sqrt{h}\, d^{3}x+\intop_{\tilde{M}}\bar{\Phi}D^{M}\Phi\,\sqrt{h}\, d^{3}x
\]
the combined Einstein-Hilbert-Dirac action for the modified space-time.
The case of the graph manifold $G$ in the model (\ref{eq:minimal-model-cosmos})
representing the interactions will be shifted to a forthcoming paper.

\section{Conclusion}

In this paper we discussed the problem, how to add fermions in GR.
We showed that the boundary term of the Einstein-Hilbert action can
be interpreted as the Dirac action for a spinor field (representing
the tubular neighborhood of the boundary in the space-time). This
spinor represents a fermion in physics only for a certain class of
compact 3-manifolds, the hyperbolic knot complements. In parallel
we discussed a family of topologically trivial space-time, the exotic
$\mathbb{R}^{4}$, which contains non-trivial compact 3-manifolds
naturally. Here one uses a property of all known exotic $\mathbb{R}^{4}$:
every compact 4-dimensional submanifold is surrounded by a neighborhood
with boundary a non-trivial compact 3-manifold (not homeomorphic to
the 3-sphere). Then the standard $\mathbb{R}^{4}$ as space-time with
no matter content is changed to a theory containing matter for the
exotic $\mathbb{R}^{4}$.

This work is an extension of the work \cite{AsselmeyerRose2012} for
the compact case (using Fintushel-Stern knot surgery) to the non-compact
case, in particular the exotic $\mathbb{R}^{4}$. The technique is
different from \cite{AsselmeyerRose2012}. Here we used a general
embedding of the 3-manifold into the 4-manifold to construct the Dirac
action. The construction of a special surfaces as well the Weierstrass
representation is obsolete and not needed anymore. Secondly, we derived
the Dirac action from the boundary term in the Einstein-Hilbert action
but discussed also all other properties like spin and state equation
in cosmology. Then our method can be reversed: one can add fermions
to a free, diffeomorphism-invariant theory containing the Einstein-Hilbert
action by adding a non-trivial 3-manifold to the spatial component
of the space-time. The theory in this paper also differs in the physical
interpretation from the previous work \cite{AsselmeyerRose2012}.
Now the fermion field is given by the embedding of the 3-manifold
and a particular realization of this embedding (excitation of the
field) describes a fermion (hyperbolic knot complement). The embedding
determines the dynamics (Dirac equation) but is independent of the
3-manifold. If we are able to interpret the hyperbolic knot complements
as fermions then we obtained also a theory which has a non-constant
number of particles. This behavior is known from QFT. Furthermore,
our theory contains also link complements of a fixed type, the Whitehead
link complement. We conjecture that this complement describes the
cloud of virtual particles (mainly virtual fermions). The particular
choice of the neighborhood determines the complexity of the 3-manifold
and also the number of fermions (here the number of hyperbolic knot
or link complements). Certainly more work is needed to understand
this property and also the inclusion of interactions more fully. At
the end we will mention one property which is independent of exotic
smoothness: \emph{Adding a hyperbolic knot complement to the 3-manifold
$\Sigma$ (representing the space) is the same as adding a Dirac action
of the space-time $\Sigma\times[0,\epsilon)$ (for suitable values
of $\epsilon$)}.

\section*{Acknowledgment}

This work was partly supported (T.A.) by the LASPACE grant. The authors
acknowledged for all mathematical discussions with Duane Randall,
Robert Gompf and Terry Lawson. Furthermore we thank the two anonymous
referees for pointing out some errors and limitations in a previous
version as well for all helpful remarks to increase the readability
of the paper.

\appendix

\section{Connected and boundary-connected sum of manifolds\label{sec:Connected-and-boundary-connected}}

Now we will define the connected sum $\#$ and the boundary connected
sum $\natural$ of manifolds. Let $M,N$ be two $n$-manifolds with
boundaries $\partial M,\partial N$. The \emph{connected sum} $M\#N$
is the procedure of cutting out a disk $D^{n}$ from the interior
$int(M)\setminus D^{n}$ and $int(N)\setminus D^{n}$ with the boundaries
$S^{n-1}\sqcup\partial M$ and $S^{n-1}\sqcup\partial N$, respectively,
and gluing them together along the common boundary component $S^{n-1}$.
The boundary $\partial(M\#N)=\partial M\sqcup\partial N$ is the disjoint
sum of the boundaries $\partial M,\partial N$. The \emph{boundary
connected sum} $M\natural N$ is the procedure of cutting out a disk
$D^{n-1}$ from the boundary $\partial M\setminus D^{n-1}$ and $\partial N\setminus D^{n-1}$
and gluing them together along $S^{n-2}$ of the boundary. Then the
boundary of this sum $M\natural N$ is the connected sum $\partial(M\natural N)=\partial M\#\partial N$
of the boundaries $\partial M,\partial N$.

\section{Spin $\frac{1}{2}$ from space a la Friedman and Sorkin\label{sec:Spin-from-space}}

As shown by Friedman and Sorkin \cite{FriedmanSorkin1980}, the calculation
of the angular momentum in the ADM formalism is connected to special
diffeomorphisms $R(\theta)$ (rotation parallel to the boundary w.r.t.
the angle $\theta$). So, one can speak of spin $\frac{1}{2}$, in
case of $R(2\pi)\not=-1$. Interestingly, these diffeomorphisms are
well-defined on all hyperbolic 3-manifolds.

In the following we made use of the work \cite{FriedmanSorkin1980}
in the definition of the angular momentum in ADM formalism. In this
formalism, one has the 3-manifold $\Sigma$ together with a time-like
foliation of the 4-manifold $\Sigma\times\mathbb{R}$. For simplicity,
we consider the interior of the 3-manifold or we assume a 3-manifold
without boundary. The configuration space $\mathcal{M}$ in the ADM
formalism is the space of all Riemannian metrics of $\Sigma$ modulo
diffeomorphisms. On this space we define the linear functional $\psi:\mathcal{M}\to\mathbb{C}$
calling it a state. In case of a many-component object like a spinor
one has the state $\psi:\mathcal{M}\to\mathbb{C}^{n}$. Let $g_{ab}$
be a metric on $\Sigma$ and we define the generalized position operator
\[
\hat{g}_{ab}\psi(g)=g_{ab}\psi(g)
\]
together with the conjugated momentum 
\[
\hat{\pi}^{ab}\psi(g)=-i\frac{\delta}{\delta g_{ab}}\psi(g)\quad.
\]
Let $\phi_{\alpha}$ with $\alpha=1,2,3$ be vector fields fulfilling
the commutator rules $[\phi_{\alpha},\phi_{\beta}]=-\epsilon_{\alpha\beta\gamma}\phi_{\gamma}$
generating an isometric realization of the $SO(3)$ group on the 3-manifold
$\Sigma$. The angular momentum corresponding to the initial point
$(g_{ab},\pi^{ab})$ with the conjugated momentum $\pi^{ab}=(16\pi)^{-1}(-K^{ab}+g^{ab}K)\sqrt{g}$
(in the ADM formalism) and the extrinsic curvature $K_{ab}$ is given
by 
\[
J_{\alpha}=-\int\limits _{\Sigma}\mathcal{L}_{\phi_{\alpha}}(g_{ab})\pi^{ab}\: d^{3}x
\]
with the Lie derivative $\mathcal{L}_{\phi_{\alpha}}$ along $\phi_{\alpha}$.
The action of the corresponding operator $\hat{J}_{\alpha}$ on the
state $\psi(g)$ can be calculated to be 
\[
\hat{J}_{\alpha}\psi(g)=-i\frac{d}{d\theta}\psi\circ R_{\alpha}(\theta)^{*}(g)|_{\theta=0}
\]
where $R_{\alpha}(\theta)$ is a 1-parameter subgroup of diffeomorphisms
generated by $\phi_{\alpha}$. Then a rotation will be generated by
\[
\exp(2\pi i\hat{J})\psi=\psi\circ R(2\pi)^{*}\qquad.
\]
Now a state $\psi$ carries spin $\frac{1}{2}$ iff $\psi\circ R(2\pi)^{*}=-\psi$
or $R(2\pi)=-1$. In this case the diffeomorphism $R(2\pi)$ is not
located in the component of the diffeomorphism group which is connected
to the identity (or equally it is not generated by coordinate transformations).

\section{Scalar curvature and energy density\label{sec:Scalar-curvature-and-energy-density}}

Let us consider a Friedmann-Robertson-Walker-metric 
\[
ds^{2}=dt^{2}-a(t)^{2}h_{ik}dx^{i}dx^{k}
\]
on $N\times[0,1]$ with metric $h_{ik}$ on $N$ and the Friedmann
equation

\[
\left(\frac{\dot{a}(t)}{c\cdot a(t)}\right)^{2}+\frac{k}{a(t)^{2}}=\kappa\frac{\rho}{3}
\]
with the scaling factor $a(t)$, curvature $k=0,\pm1$ and $\kappa=\frac{8\pi G}{c^{2}}$.
As an example we consider a 3-dimensional submanifold $N$ with energy
density $\rho_{N}$ and curvature $R_{N}$ (related to $h$) fixed
embedded in the space-time. Next we assume that the 3-manifold $N$
possesses a homogenous metric of constant curvature. For a fixed time
$t$, the scalar curvature of $N$ is proportional to 
\[
R_{N}\sim\frac{3k}{a(t)^{2}}
\]
and by using the Friedmann equation above, one obtains 
\[
\rho_{N}=\frac{1}{\kappa}R_{N}+\rho_{c}
\]
with the critical density 
\[
\rho_{c}=\frac{3}{\kappa}\left(\frac{\dot{a}(t)}{c\cdot a(t)}\right)^{2}=\frac{3H^{2}}{\kappa}
\]
and the Hubble constant $H$ 
\[
H=\frac{\dot{a}}{c\cdot a}\quad.
\]
The total energy of $N$ is given by 
\begin{equation}
E_{N}=\intop_{N}\rho_{N}\,\sqrt{h}d^{3}x=\frac{1}{\kappa}\intop_{N}R_{N}\sqrt{h}d^{3}x+\rho_{c}\cdot vol(N)\,.\label{eq:total-energy}
\end{equation}
For a space with constant curvature $R_{N}$ we obtain 
\begin{equation}
E_{N}=\left(\frac{1}{\kappa}R_{N}+\rho_{c}\right)\cdot vol(N)\label{eq:total-energy-constant-curvature}
\end{equation}


\end{document}